# Ultrafast interband transitions in nanoporous gold metamaterial

Tlek Tapani[1], Jonas M. Pettersson[1], Nils Henriksson[1], Erik Zäll[1], Nils V. Hauff[2], Lakshmi Das[1], Gianluca Balestra[3,4], Massimo Cuscunà[3], Aitor De Andres[1], Tommaso Giovannini[5], Denis Garoli[6,7], and Nicolò Maccaferri[1]*

1. Department of Physics, Umeå University, Umeå, Sweden
2. Umeå Centre for Electron Microscopy, Umeå University, Umeå, Sweden
3. CNR NANOTEC Institute of Nanotechnology, Via per Monteroni, Lecce, Italy
4. Department of Mathematics and Physics, University of Salento, Via per Monteroni, Lecce, Italy
5. Department of Physics, University of Rome Tor Vergata, Rome, Italy
6. Dipartimento di Scienze e Metodi dell'ingegneria, Università di Modena e Reggio Emilia, Reggio Emilia, Italy
7. Istituto Italiano di Tecnologia, Genova, Italy
*nicolo.maccaferri@umu.se

**Nanoporous metals have emerged as promising functional architectures due to their tunable optical and electronic properties, high surface areas, and versatile use in real-life applications such as sensing, catalysis, and biomedicine. While the optical and morphological properties of nanoporous metals have been extensively studied, their electronic properties at ultrafast timescales remain largely unexplored. Here, we study the transient response of a nanoporous gold metamaterial and compare it with the ultrafast dynamics of a continuous gold film. We unravel that the nanoporous sample supports lower energy interband transitions, due to a much higher electron temperature in the nanoporous material, which causes an enhanced redistribution of electron density around the Fermi level. The experimental results are consistent with the two-temperature model, which highlights the role of nanoscale porosity in enabling the more efficient generation of hot carriers, thus allowing lower energy photons to induce interband transitions. Our findings demonstrate that nanoporosity affects fundamental ultrafast electronic processes and introduces this platform as temporal metamaterial allowing the emergence of tunable electronic properties not supported by the bulk counterpart. Furthermore, we present new insights into ultrafast electronic properties of nanoporous metals, which can impact several areas, from photochemistry and catalysis to energy harvesting and opto-electronics.**

**Introduction**

Nanoporous materials are known to be extremely efficient in enhancing light-matter interactions[1], enabling applications in nanophotonics[2,3], biomedicine[4–8], spintronics[9], and energy conversion[10]. In this context, nanoporous gold (NPG) films have emerged as promising systems due to their tunable optical properties[11], high surface area and interconnected nanostructures, making them highly versatile for applications in sensing[12–14] and catalysis[15,16]. Furthermore, the presence of nanoscale pore-like structures and grain boundaries introduces additional degrees of freedom for tuning plasmon-induced phenomena[17,18], making NPG an attractive platform for developing efficient hot-electron generators[19] and optical filters[20,21], where controlled optical absorption is crucial. It is well known that gold (Au) exhibits interband transitions in the visible spectral range close to 2.4 eV (~ 520 nm) due to transitions from the 5d band to the conduction 6sp band[22]. In the context of plasmonics, while previous research focused primarily on how interband transitions influence plasmonic behavior[23–25], the cross talking between plasmons and interband transitions remains less explored, in particular at ultrafast (femto- to picosecond) timescales[26]. Only recently, studies have pointed out the role of interband and plasmonic excitations on the ultrafast carriers dynamics[27–29]. Indeed, while interband transitions are typically regarded as loss channels in plasmonic systems[30], their potential active role in ultrafast carrier generation remains an open question[31–33]. Here, we experimentally and theoretically unveil how the nanoporosity of an archetypical plasmonic material, such as gold, can induce transient interband transitions below 2.3 eV (~ 540 nm), which is the typical energy at which these electronic transitions are observed in bulk gold (BG) in pump-probe experiments[34]. By combining ultrafast transient transmission, cathodoluminiscence (CL) and linear absorption spectroscopy we show how the generation of hot carriers in NPG can provide more empty states levels in the 6sp conduction band, consequently requiring lower photon energy for interband electron excitation from the 5d band compared to a continuous gold film. This has direct implications for hot carrier generation, as the increased availability of low-energy empty states in the conduction band facilitates more efficient excitations and relaxation pathways. Therefore, nanoporosity effectively broadens the spectral window and temporal dynamics of hot carrier populations, questioning the common notion that interband transitions are only passive loss channels[35]. To support our findings, we use an analytical approach based on the two-temperature model (2TM) and the transfer matrix method (TMM) to assess the role of nanoporosity in affecting transient low energy interband transitions in NPG. Our results provide new fundamental insights into how plasmonic and interband excitations synergistically combine at the ultrafast level, and how nanoscale porosity affects electron dynamics and influences the energy dispersion in noble metals at ultrafast timescales. Such understanding paves the way for developing advanced plasmonic platforms based on NPG where control over carrier generation and relaxation is essential, including in hot carrier driven processes, as for instance light harvesting and catalysis[36,37].

**Results**

NPG films were synthetized on fused silica substrates using a recently developed dry method[38,2] (see Methods for more details). This approach enables the preparation of NPG thin films with no impurities from other metals, a typical limitation observed in the standard preparation techniques based on dealloying[1]. As reference sample, a bare gold film thermally evaporated on glass was used. X-ray photoelectron spectroscopy (XPS) measurements were performed on the samples to ensure that no oxidation occurred, in particular on the NPG films (see Supplementary Note 1).

We performed pump-probe measurements (for more details see Methods section) to understand the ultrafast carriers' dynamics upon photoexcitation with sub-10 fs light pump pulse (central wavelength 850 nm, 1.46 eV), and we probed the dynamics using a broadband pulse, covering a spectral range from 500 to 750 nm (pulse duration sub-15 fs). In Figure 1, a sketch of the different electronic transitions probed by our light pulses are reported. In both BG and NPG films, after pump excitation, intraband transitions from electronic states around the Fermi level $E_f$ and the empty states above (red arrows in the left and right panels) can take place. When the photon energy is larger, interband transitions from the 5d band to the 6sp band can be promoted. In BG, electronic transitions typically require energies exceeding 2.3 eV (green arrow in Figure 1a). In contrast, in NPG, the elevated hot electron temperatures significantly alter the Fermi-Dirac distribution near the $E_f$. This results in a greater availability of empty states in the 6sp band at lower energies, allowing lower-energy probe photons to induce these transitions (orange arrow in Figure 1b). This can be explained by using the 2TM, similar to the one developed previously by Ortolani et al.[19], which allows us to estimate the effective fluence acting on the carriers upon photoexcitation depending on the geometry of the sample and the filling factor, and thus determine the electronic temperature in both samples (see Methods for more details).

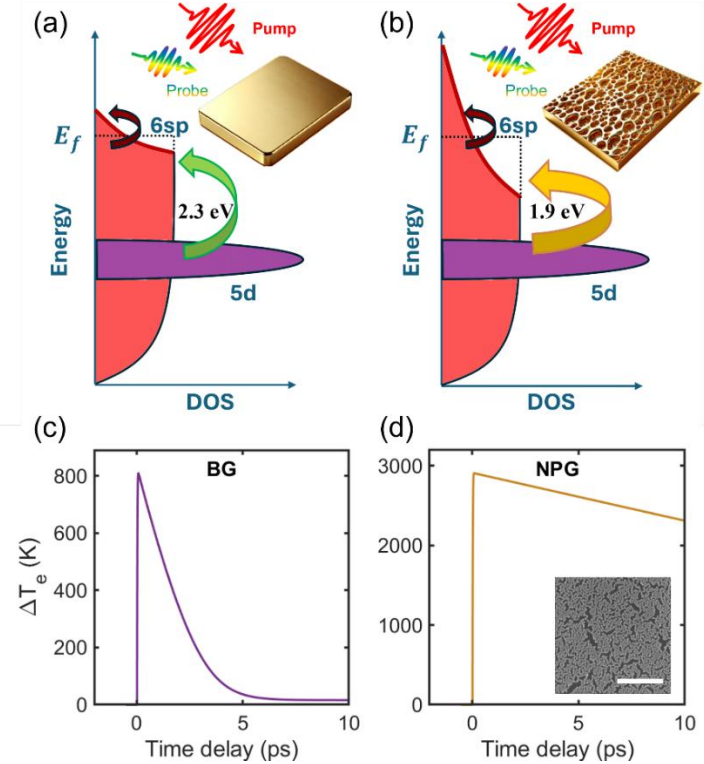

**Figure 1. Ultrafast interband and intraband transitions in bulk and nanoporous Au.** (a) In BG, the interband transition threshold (~2.3 eV, 540 nm) corresponds to transitions from the 5d band to the 6sp conduction band. (b) In NPG the pump pulse alters the local density of electronic states in a stronger way due to the enhanced light-matter coupling, leading to a stronger redistribution of transition probabilities. The shift to a lower-energy transition (~650 nm, 1.9 eV) indicates a reduced "effective" bandgap. (c-d) Transient variation of the electronic temperature in BG (c) and NPG (d) as computed by using the 2TM (see Methods). Inset in (d) is a SEM image of the NPG. Scale bar 300nm.

In our model, we assume that both systems are excited by a pulse centered at 850 nm and a fluence of 3 mJ/cm$^2$ closely matching our experimental parameters. In Figure 1c and 1d, the maximum variation of the electronic gas temperature in the BG and NPG is ~800 K and ~2900 K, respectively. These temperature variations correspond to roughly 0.07 eV for BG and 0.25 eV for the NPG film. This means that, in the latter case, more empty states closed to the 5d band become available, and thus probe photons at lower energies (below 2.3 eV) can promote interband transitions, as depicted in Figure 1b. This effect is triggered directly by the higher absorption in NPG film at the pump energy (see Supplementary Note 2). Interestingly, it can be noticed from the simulation that, while BG present a quite sharp exponential decay of the electronic temperature upon light excitation, which is typically expected for these laser fluences, the NPG film displays a much longer decay time.

In addition, the larger absorption in the range 550-700 nm in the NPG film is also increasing the probability of probe photon absorption, thus promoting more efficiently interband transitions in that spectral range. This higher absorption can be related to the presence of localized plasmonic resonances, which extend over a broad spectral range due to the high dispersion in size and shape of the nanoporous material[1]. To investigate this, we performed CL spectroscopy (for more details see Methods section). In Figure 2a, a scanning electron microscopy (SEM) image of a portion of the NPG film is shown. It can be observed that NPG film's interconnecting wire size is in the similar range of the mean free path of Au electrons, that is around 40 nm[39]. This leads to an increased surface scattering which reduces electron-phonon coupling strength in the porous structure compared to the BG film case, and thus explaining the longer relaxation time observed in Figure 1d. The maximum CL signal map in Figure 2b shows variations in localized surface plasmon resonance wavelengths across the sample due to the high anisotropy and dispersion in size of the metallic clusters in NPG films. Additionally, while the hole and gaps emit in the blue part of the spectrum, the regions containing Au emit more in the red part of spectrum, around between 550 and 700 nm.

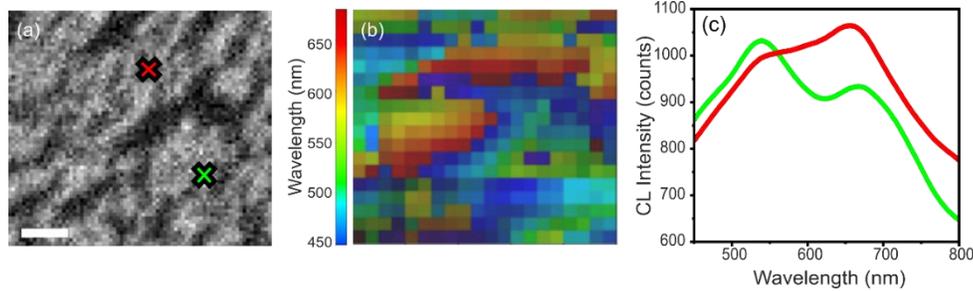

**Figure 2. Cathodoluminescence spectroscopy measurements on NPG film.** (a) SEM image of nanoporous Au; scale bar: 100 nm. (b) Wavelength of maximum CL signal from the portion of the sample shown in (a). (c) CL spectra taken at two different locations highlighted in (a), revealing the presence of localized plasmons resonances in the range 500 – 700 nm.

Figure 2c confirms this spatially dependent emission behavior, displaying the CL spectrum of the points marked in Figure 2a, which reveals a broad localized plasmon resonance centered around 600 nm. The emission is three times larger in terms of counts compared to the previously mentioned hole and gap regions emitting in the blue. The BG film instead show a clear peak close to 540 nm (2.3 eV), which is exactly the onset of the interband transition from 6d to 5sp in BG (see Supplementary Note 3).

We then measured the ultrafast optical response of both NPG and BG films using pump-probe spectroscopy. Transient transmission ΔT/T measurements showing the evolution of the signal over the time delay between pump and probe pulses and across different wavelengths (500 – 750 nm) are plotted in Figure 3a and 3b for NPG and BG films, respectively. The NPG film shows a peculiar transient transmission response characterized by a broad negative signal spanning from 550 nm to 670 nm. This feature is absent in the continuous gold film, where a negative signal is present only at wavelengths shorter than 550 nm, consistent with the presence of interband transitions, and disappears at longer wavelengths, as expected from previous studies on BG films[34,40]. As sketched in Figure 1b, our experimental results suggest that in the NPG film there is an enhanced redistribution of the electronic states around the Fermi level which allows probe light absorption to induce lower energy interband transitions. This phenomenon can be attributed to two main mechanisms: (i) the larger absorption and consequent higher electronic temperature in NPG at the pump pulse energy, that modify the electronic environment and redistribute electron density around the Fermi level much more strongly than in the BG film; (ii) the probe-induced excitation of localized plasmons, which contribute to a higher probe photon absorption probability and consequent interband transition in a range of energies below the BG interband transition energy. The positive (red) signal in both samples reflects the fact that we are in the presence of intraband transitions, where due to bleaching of excited states in the 6sp band the probe photons are not absorbed and thus transmission increases. The yellow and purple cut lines in the maps in Figure 3a and 3b correspond to the ΔT/T signals of NPG and BG films, respectively, in the time domain at a fixed wavelength in Figure 3c. The comparison is done at 720 nm where the pump pulse mainly excites electron in the conduction band (promoting mainly intraband transitions). First, a larger positive ΔT/T value is observed, indicating a much higher electron temperature in the porous sample, in line withour calculations using the 2TM (see Figure 1c, BG, and 1d, NPG). Also, longer relaxation dynamics can be observed in the NPG film case, as already estimated in Figure 1d using the 2TM considering the geometry (filling factor) of the sample. This result can be explained by considering that in NPG film, which is characterized by significantempty spaces, the thermal capacitance is reduced and thus the electron-phonon coupling decreases[41]. In other words, the presence of holes and gaps in the NPG film limits heat transfer between electrons and phonons. This results also in a higher electron temperature, and consequently a larger variation of the dielectric permittivity of the sample, thus of the transient transmission signal. Electron thermalization occurs through both electron-electron and electron-phonon scattering, but here we mainly focus on the second process as the first one is very fast (few tens of fs) in Au. Electron-phonon coupling is quite different if we compare BG and NPG films. In the latter such a coupling is weaker, and this increases the relaxation time[19]. In our case, this effect drastically affects the decay of the pump-probe signal for the BG compared to the NPG film case.

We now focus on the spectral dependence of the transient signal of our samples. The pink and green cut lines in the maps in Figure 3a and 3b correspond to the ΔT/T signals of NPG and BG films, respectively, in the frequency domain at a fixed time delay (400 fs), as shown in Figure 3d. The BG film exhibits a strong negative signal around 520 nm, indicating increased absorption due to interband transitions. This result is consistent with the theory and the well-known ultrafast dynamics of Au[34,40]. On the contrary, the NPG film displays a broader negative signal, suggesting a modified interband transition response, as also sketched in Figure 1b. The differences between the two curves highlight the impact of the nanoporous structure on the transient electronic properties, particularly in how it influences interband transitions in the time domain.

To support these hypotheses based on our experiments, we employed a 2TM model taking into account the geometry of the NPG, in combination with the TMM to calculate the ΔT/T signal (see Methods for more details).Our model does not consider the presence of plasmonic excitations in NPG, because we wanted to focus on the nanoporous geometry role in affecting interband transitions and that broadband plasmonic excitations are merely contributing to enhance probe absorption, but they do not affect the physics of the mechanism investigated here, which is mainly connected to the porosity of the system[35,42].

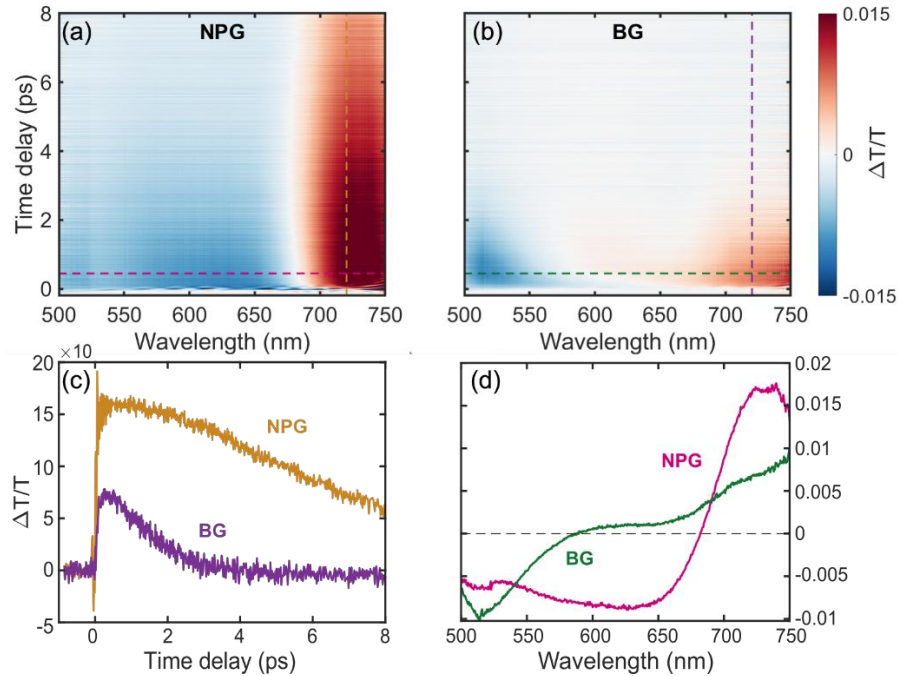

**Figure 3. Transient optical spectroscopy experiments of nanoporous and bulk gold films.** ΔT/T of NPG (a) and BG (b) films, as a function of time delay between pump and probe pulses and wavelength of the probe light pulse. (c) ΔT/T of nanoporous gold (yellow curve) and bulk gold (purple curve) films as a function of the time delay between pump and probe pulses at 720 nm (intraband transitions spectral region). (d) ΔT/T of NPG (pink curve) and BG (green curve) films as a function of the probe light wavelength and at a time delay of 400 fs.

In Figure 4a and 4b we plot the calculated ΔT/T signal as a function of the pump-probe time delay and the wavelength of the probe pulse wavelength for the NPG and BG film, respectively. The NPG film displays a much longer time decay compared to the BG film. Moreover, while the negative signal in the bulk gold film is centered at the expected interband transition energy, the calculated response in the NPG film is broader. Thus, the modelreproduces the observed experimental results to a relatively high extent. To further support our hypothesis that the nanoporosity of our sample is indeed the reason behind this low energy and broader interband transitions response, we calculated the ΔT/T signal as a function of the probe light wavelength and the filling factor. Figure 4c shows that by increasing the filling factor, the negative signal becomes broader, or, in other words, the crossing point from negative to positive signal red shifts. SEM images analysis indicates a filling factor of 0.43 with our sample (dashed line in Figure 4c). Since we cannot have the same control over the filling factor in our experiments, we tried to find an alternative way to support that porosity indeed affects the probability for interband transitions below 2.3 eV. One way to do so is to increase the fluence of the pump pulse. Increasing the pump fluence leads to an increased electronic temperature in the NPG film, thus making more states available in the 6sp band at lower energies.

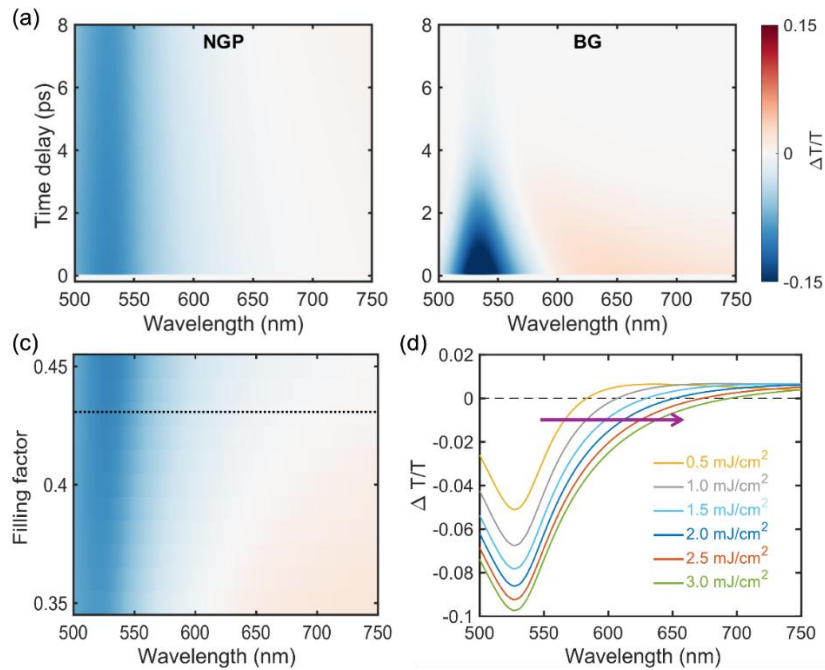

**Figure 4. Simulations of the ultrafast response of nanoporous and bulk gold.** Calculated ΔT/T of NPG (a) and BG (b) films, as a function of time delay between pump and probe pulses and wavelength of the probe light pulse using 2TM and TMM. (c) Calculated ΔT/T of NPG as a function of the wavelength of the probe light pulse and the filling factor at time delay of 100 fs, the color bar scale same as in (a-b); the dashed line highlights the filling factor of the NPG sample measured experimentally in (a), that is f = 0.43. (d) Calculated ΔT/T of NPG as a function of the wavelength of the probe light pulse at time delay of 100 fs at different pump fluences; the purple arrow highlights the shift of the crossing point between negative and positive signal, and thus the broadening of the interband transition band induced by the NPG film porosity.

Thus, one can expect a red-shift of the crossing point between negative-to-positive signals by increasing the pump fluence. This is indeed what we observe in Figure 4d (see red arrow pointing towards larger wavelengths as the fluence of the pump pulse increases), where we plot the ΔT/T signal as a function of the probe light wavelength for different pump pulse fluences. It is worth noting here that our model does not account for higher order effects which might impact the saturation of the effect observed in Figure 4d. In our 2TM we consider only the linear absorption of our sample to estimate the impact of the power source term in our model. Nevertheless, the theory captures qualitatively very well the experimental observations reported in Figure 3, thus confirming that porosity plays a significant role in affecting the transient behavior of interband transitions in NPG films.

To confirm the prediction reported in Figure 4d, we performed fluence dependence measurements on the NPG film, where we studied the variation of the ΔT/T signal at different pump light powers. In Figure 5a we plot the ΔT/T signal of the NPG film as a function of the probe light wavelength and at a time delay of 440 fs, measured at pump pulse fluences spanning from 2 to 3.5 mJ/cm$^2$. As it can be seen, the absolute intensity of the negative signal due to interband transitions increases by increasing the pump fluence, and signal overall undergoes a red-shift up to 3 mJ/cm$^2$. This is consistent with the picture predicted by our model, where lower fluence means less energy in the system, which in turn leads to a smaller modification of the Fermi-Dirac distribution around $E_f$ and thus a smaller probability to have interband transitions at probe photon energies below 2.3 eV. Figure 5b shows that the minimum of ΔT/T for NPG film red-shifts. This is in line with the red-shift of zero-crossing point in the simulations (see Figure 4d). Noteworthy, ΔT/T for BG film exhibits a negligible shift (see Supplementary Note S4). Thus, we can claim that the stronger response of the NPG film arises mainly from geometric confinement, as accounted in our simulations. Nevertheless, although not taken into account in our calculations, we can also hypothesise that plasmonic effects are also contributing to promote the low energy interband transitions not present in the BG film. It is worth mentioning here that, in the experiment, the zero-crossing wavelength is not used as a reference point as in the simulation due to the fact that this point is very close to the pump central wavelength, and it is affected by the ground (6sp) state bleaching[43], that is the strong positive signal between 700 and 750 nm, which is not present in our calculations. This positive signal affects more strongly the zero-crossing value than the ΔT/T minimum. We also see that above 3.5 mJ/cm$^2$ the effect reaches its saturation, which is also visible in our modelling. This can be explained by the fact that after a fluence of 3 mJ/cm$^2$ the dependence of the pump-probe signal on the pump light power starts to be nonlinear (see Supplementary Note S5) both in the intraband region, where we pump the system, but also in the interband region where it starts to diminish as we increase the pump fluence.

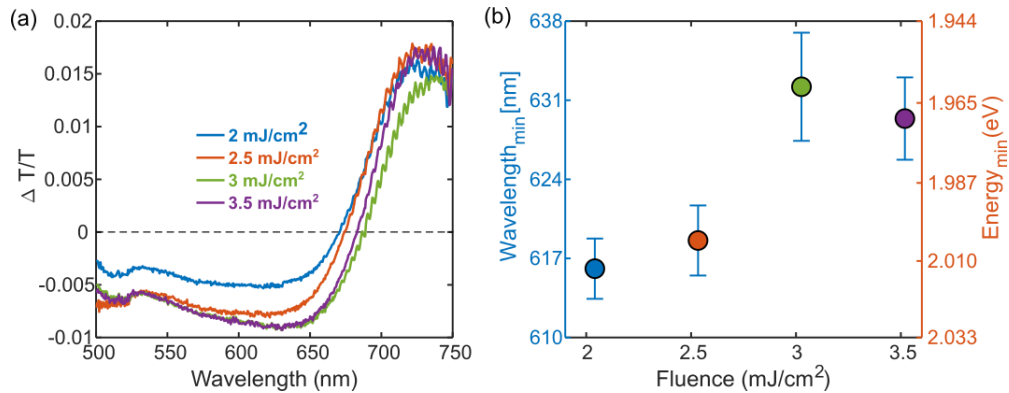

**Figure 5. Power dependence of transient transmission.** (a) Pump fluence dependence of ΔT/T spectrum at a time delay of 440 fs. A red-shift in the minimum of the ΔT/T signal, as well as of the crossing point between negative-to-positive signal, is observed with increasing pump intensity. (b) Wavelength of ΔT/T signal minimum as a function of the pump pulse fluence.

**Discussion**

In this work we report the experimental observation of transient low energy interband transitions in NPG metamaterial thin film by using transient transmission pump-probe spectroscopy. While our benchmark BG film presents the well-known interband transitions around 2.3 eV, the NPG film display a broader response which suggest that, due to porosity, interband transitions are allowed at lower energies upon probe light absorption. In the same spectral region where absorption is already high due to the presence of broadband plasmonic resonancebelow 2.4 eV, which has been proved by using CL spectroscopy, the combined effect of porosity and plasmonic modes further amplifies the probe light absorption and promote this broader low energy interband transitions. To support our observations, we developed an analytical approach based on the 2TM and the TMM, which consider the geometry of the porous material. In our model, we found that porosity is indeed crucially affecting the interband transitions, and that the filling factor, thus the geometry, of the nanoporous material has a relevant effect in the broadening of the transient low energy interband transitions. This is mainly due to the fact that, by increasing the filling factor, larger absorption of light is possible at the pump wavelength. This in turn means that we reach a higher electronic temperature in the system, thus promoting intraband transitions to higher levels and thus leaving more empty states at lower energies in the 6sp band, allowing interband transitions from the 5d band with photons of energy below 2.3 eV. We prove this last observation by performing fluence dependence measurements, which points in the same direction as suggested by our model. Overall, our study presents a new concept of how porosity in metals can reshape electronic transitions at ultrafast time scales. By decoupling the role of geometry from plasmonic effects, we show that porosity alone can shift and broaden interband features, offering an additional degree of control over light-matter interactions. This insight opens new opportunities in fields where nanoporous materials are already central, such as plasmon-enhanced catalysis, ultrafast photochemistry, and active nanophotonics.

**Methods**
Sample fabrication. NPG thin films were prepared following a dry process method reported in Refs. [2,38]. In summary, a Poly(methyl methacrylate) (PMMA) thin layer was spin-coated on a glass substrate and baked at 180 °C for 3 minutes. Then, a thin layer of Au was deposited by means of electron-beam evaporation using a tilted configuration (i.e. placing the sample tilted at 80° with respect to the evaporation source). The process was completed with the etching of the PMMA layer by means of $O_2$ plasma (200 W, 10 minutes). A BG thin film was prepared using standard electron beam evaporation on a glass substrate.

Pump-probe measurements. The pump-probe setup used a Yb:KGW laser (Model: PH2-20W-SP, Light Conversion) to generate both probe and pump pulses. The probe was generated through a 6-mm-thick YAG crystal (undoped YAG, orientation [100]), and a compressor composed of dielectric chirped mirrors (custom made from Ultrafast Innovation). This produces pulses with a spectrum ranging from 1.3 – 2.5eV (white light) with sub-15 fs duration and repetition rate of 50 kHz. The pump was generated by using a Visible to Near-Infrared Optical Parametric Amplifier (VIS-NIR OPA, Model: ORPHEUS-N-2H, Light Conversion), producing a spectrum ranging from 1.4-1.7eV with 9 fs duration. A tunable intensity filter was also added for power dependence measurements, an external pulse picker was adopted to halve the repetition rate of pump to 25kHz for the differential transmission (ΔT/T) measurements. A spectrograph (Model: HSVIS camera with sensor HA:S14290Q-1024) was used to collect the transmitted probe that provides two-dimensional ΔT/T maps as a function of pump-probe delay t and probe wavelength λ.

CL measurements. Nanoporous Au films were fabricated on a 100 nm thick $Si_3N_4$ membrane to significantly minimize emission from bulky substrates during cathodoluminescence (CL) investigation. CL analysis was performed at room temperature by using a Zeiss Merlin scanning electron microscope (SEM) equipped with a high-performance CL imaging system (SPARC from Delmic). CL was spectrally resolved in the range of 450-800 nm with an "Andor Kimera 193i" spectrometer with a focal length of 193 mm and a grating of 300 gr/mm. The photon emission was captured by an "Andor Newton DU920P-BEX2DD" CCD camera with a maximum quantum efficiency of 90%. The electron beam operated at an acceleration voltage of 30 kV and an emission current of 10 nA. The color-coded map consisted of 20 × 18 pixels (pixel size: 25 nm). The focused electron beam was scanned across the specimen, dwelling for 10 s (integration time) on each pixel to acquire the CL spectra.

Modelling. The full simulation of a pump-probe experiment is here done in three steps: first by calculating the pump-matter interaction, followed by calculation of the transient change in permittivity induced by the pump, and lastly the probe interaction with the excited structure. The optical response of the NPG and BG thin films was calculated using the transfer matrix method using the Drude-Lorenz model for the Au permittivity by Rakić et al.[44], with the NPG represented by a thin film with an effective permittivity[45]. Due to the high filling fraction of Au in the NPG, the Bruggeman effective medium model was employed over the Maxwell-Garnett model[46,47]. The absorption $A$ of the NPG and BG at the pump wavelength of 850 nm was then used as input into a two-temperature model[48],

$$C_e \frac{\partial T_e}{\partial t} = -G(T_e)(T_e - T_l) + S(t)$$
$$C_l \frac{\partial T_l}{\partial t} = G(T_e)(T_e - T_l)$$

where $T_{e(l)}$ is the electron (lattice) temperature, $G(T_e)$ is the electron-phonon coupling factor and $C_{e(l)}$ is the electron (lattice) heat capacity of Au. Both $G$ and $C_e$ are interpolated from the data work of Lin et al.[49], while $C_l = 2.49 \times 10^6$ Jkg$^{-1}$K$^{-1}$ was used for the lattice heat capacity[50]. However, for the NPG each material parameter was scaled by a factor $f_m^\beta$, where $f_m$ is the metall filling factor of 0.43 and $\beta$ a constant. For $\beta$ we used the same values as in Ref.[19] The final term $S(t)$ is the source term, corresponding to the pump pulse. In this work, a Gaussian was used to model the pulse as[51]

$$S(t) = \frac{AF}{d}\sqrt{\frac{4\ln 2}{\pi \Delta t^2}} e^{-4\ln 2 (t-t_0)^2/\Delta t^2}$$

Here, F and $\Delta t$ are the pump fluence and duration, and $d$ the NPG or BG film thickness. The probe-matter interaction at different time delays was finally calculated using the TMM with a perturbed permittivity $\varepsilon_{Au}(\lambda) + \Delta\varepsilon(\lambda, t)$, where the change in permittivity is determined from $T_e(t)$ and $T_l(t)$ and $\lambda$ is the probe wavelength. Details of this calculation can be found in other works, see Refs[32,52] and other references cited by these papers.

**Author contributions**
NM conceived the study and supervised the work. TT, JMP and ADA performed the pump-probe measurements. NH performed the calculations. DG fabricated the samples. GB and MC performed the CL measurements. EZ, JMP and LD performed linear absorption measurements. NVH and LD performed SEM characterization. TT, JMP, NH, TG and NM analyzed and discussed the results. TP and NM wrote the manuscript with contributions from all the authors.

**Acknowledgements**
This work had been funded by Swedish Research Council (Grant No. 2021-05784), the Knut and Alice Wallenberg Foundation through the Wallenberg Academy Fellows Program (Grant No. 2023.0089), the European Research Council through the ERC Starting Grant 'MagneticTWIST' (Grant No. 101116253) and the HORIZON-Pathfinder-Open '3D-BRICKS' (Grant No. 101099125). We acknowledge Dr. Andrey Shchukarev and Dr. Dmitry Shevela from the XPS Platform at the Department of Chemistry, Umeå University, for the support with XPS measurements and analysis. The authors acknowledge Umeå Centre for Electron Microscopy (UCEM) and the National Microscopy Infrastructure (NMI), for instrument access and technical support.

# SUPPLEMENTARY INFORMATION

**Table of contents**

**Supplementary Note 1: XPS measurements and analysis on BG and NPG films.** To confirm that the BG and NPG samples were not oxidized, X-ray Photoelectron Spectroscopy (XPS) was used. Binding Energy (BE) of Au 4f 7/2 photoelectron line for both BG film (84.0 eV) and NPG porous (84.1 eV) samples corresponds to gold metal[1] without any sign of oxidation. XPS survey spectra and high-resolution spectra of Au 4f 7/2 line are given in Supplementary Figure 1.

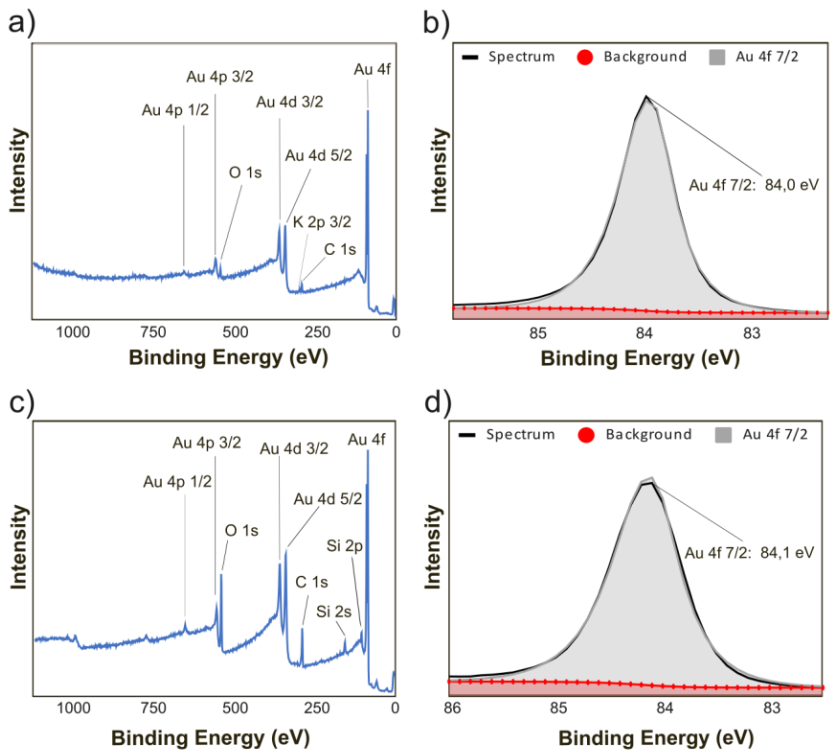

**Supplementary Figure 1**. XPS spectra for BG (a)-(b) and NPG (c)-(d) films. A90 and (c) show survey spectra with all identified elements and their photoelectron lines. (b) and (d) present corresponding high-resolution Au 4f $_{7/2}$ photoelectron lines.

**Supplementary Note 2: linear absorption measurements and simulations.** To measure the linear absorption of our samples, we used a spectrophotometer (Model: PerkinElmerLAMBDA 1050+ UV/Vis/NIR) to measure transmittance, T, and hemispherical reflectance, R. Using conservation of energy, absorption is retrieved as: A = 1-T-R. Supplementary Figure S2 compares the linear optical spectra of NPG (a) and BG (b) films (continuous lines). The dashed lines represent the calculations of the optical spectra using the TMM and the constants from Rakic [3]. The largest difference between these two is the reduction of reflectance and increase of transmittance in NPG film, which can be explained by the nanoporous structure; the porous surface causes higher scattering and the rise in transmittance is due to a smaller filling factor, decreasing the effective extinction coefficient of the film. In the BG film, the onset of interband transition from 5d band to 6sp band can be observed around 540 nm (2.3 eV).

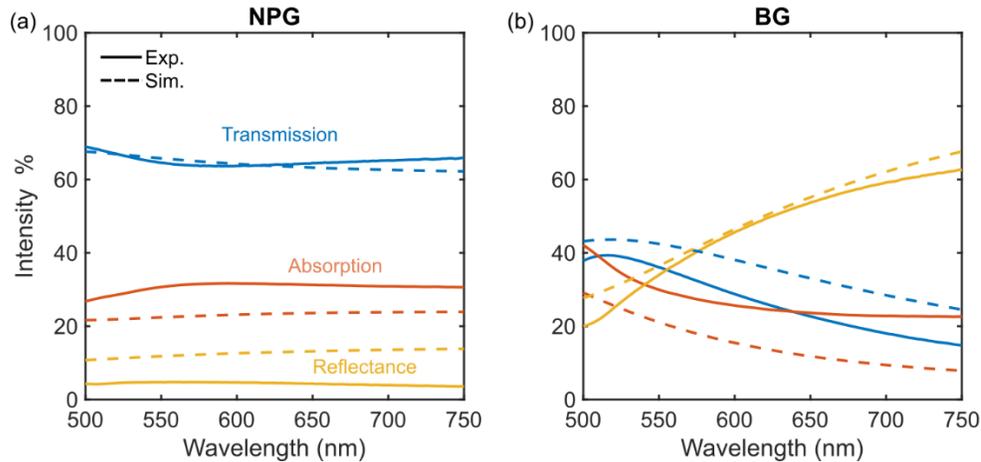

**Supplementary Figure S2.** Optical response of nanoporous (a) and bulk (b) gold films. Experimental/calculated transmittance (continuous/dashed blue lines), reflectance (continuous/dashed yellow lines) and absorption (continuous/dashed red line).

**Supplementary Note 3: cathodoluminescence spectroscopy on BG film.** To prove that the BG film does not have any plasmonic effect after the interband transition wavelength, we performed CL measurements on the BG film as well. The spectrum of each pixel shows a peak in the range 520-530 nm (interband transition of gold). Only the intensity of the peak changes but overall we do not observe additional peaks at lower wavelengths as in the case of the NPG fil (Figure 2 in the main text. The results are reported in Supplementary Figure S3.

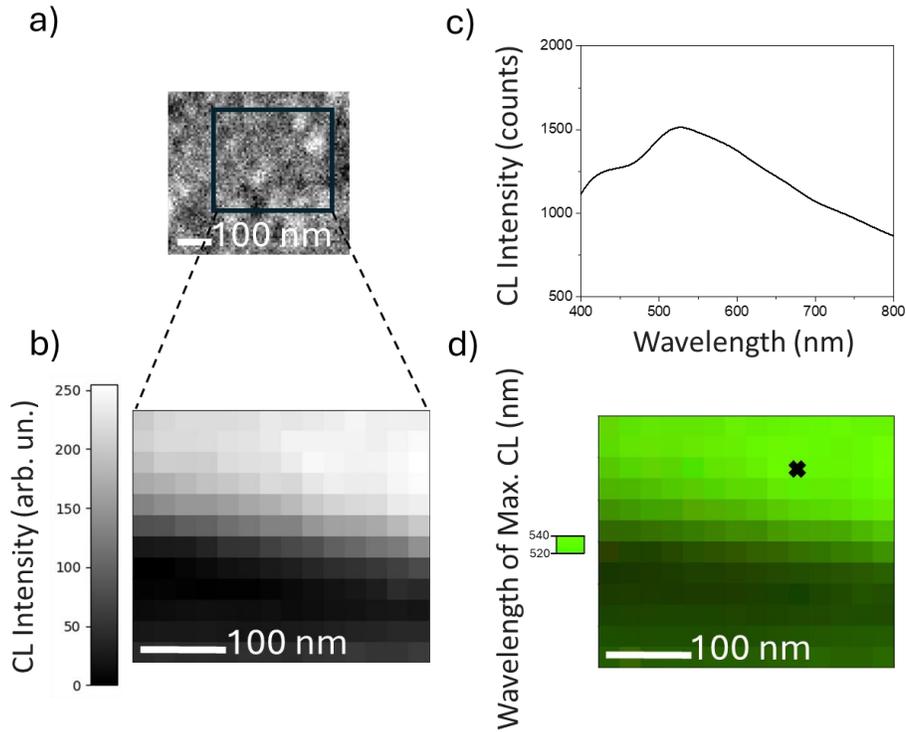

**Supplementary Figure S3.** (a) SEM image of BG film. (b) CL intensity of the area inside the black square in (a). (c) CL intensity of the pixel indicated in (d).

**Supplementary Note 4: fluence dependence of ΔT/T vs wavelength in BG film.** Unlike NPG, BG shows only a minor redshift in both simulated zero-crossing wavelength value and experimental local minimum wavelength value. Under the same pump fluence, the model predicts ≈150 nm shift of the zero-crossing wavelength for NPG but less than 50 nm for BG, as shown in Supplementary Figure S4a. Experimentally, the local spectrum shifts clearly to longer wavelengths in NPG, whereas in BG the shift is barely detectable as shown in Supplementary Figure S4b. Because the simulation includes only geometric confinement, it underestimates the NPG response: in the real sample, plasmonic hot-spots boost the hot-carrier population and drive an even larger red-shift.

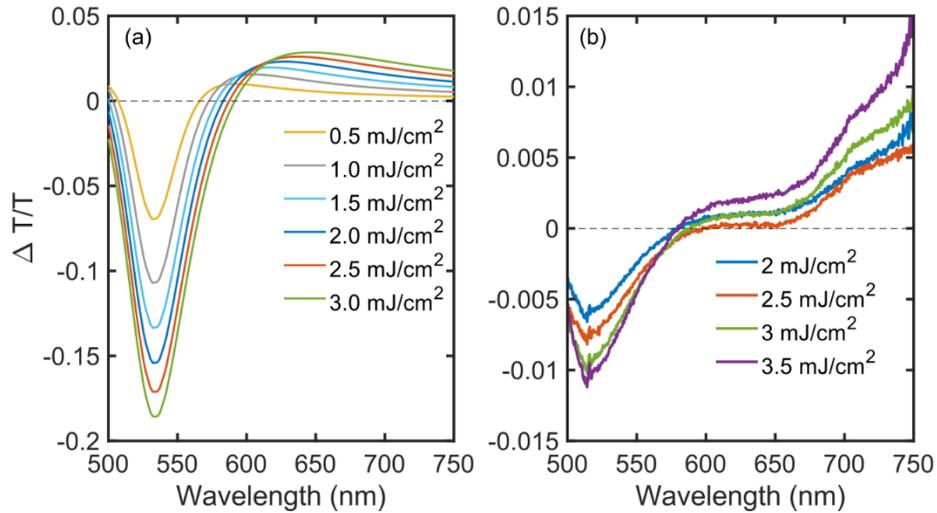

**Supplementary Figure S4. Power dependence of transient transmission in BG film.** (a) Simulated pump fluence dependence of ΔT/T spectrum at a time delay of 440 fs. A red-shift of the zero-crossing point between negative-to-positive signal is observed with increasing pump intensity. (b) Experimental pump fluence dependence of ΔT/T spectrum at a time delay of 440 fs. A red-shift in the minimum of the ΔT/T signal is observed with increasing pump intensity.

**Supplementary Note 5: fluence dependence of max(ΔT/T) and decay time in BG and NPG films**. Further investigations into the pump power-dependent response reported in Figure 5 in the main text are presented in Supplementary Figure S5. Supplementary Figure S5a and S5b show the decay time extracted by fitting the transient transmission signals of NPG and BG with an exponential decay function at fixed intraband and interband transitions, respectively. For intraband excitation at 720 nm (Supplementary Figure S5a), NPG exhibits consistently longer recovery times than the BG film across all fluences. Additionally, NPG displays a strong fluence-dependent increase in recovery time, reaching up to 10 ps at higher fluences, whereas the BG film shows negligible variation over the studied fluence range. This difference is likely due to two main factors: (i) higher steady-state absorption (~50% relative higher absorption in NPG compared to BG film, see also Supplementary Figure S2) in the pump spectral region, leading to a greater excited carrier population under the same incremental fluence, and (ii) stronger electron-phonon interactions in the BG film, enabling more efficient energy dissipation and preventing a significant increase in recovery time with fluence. From Supplementary Figure S5c it is clear that NPG exhibits higher transient transmission than the BG film, indicating a higher electron temperature. However, the change of the maximum of transient transmission as a function of pump pulse fluence for NPG is less evident. This suggests that in this fluence range, the electron heat capacity (proportional to the electron temperature) in NPG approaches a much higher value, reducing the relative change in transient signal. For interband transitions (Supplementary Figure S5b and S5d, collected at 520 nm for the BG film and at 625 nm for NPG film), both structures show a fluence-dependent increase of the recovery time and transient signal, with NPG exhibiting a steeper increase, consistent with the intraband trend. Unlike the intraband case, the maximum transient transmission signal increases similarly for both samples, which can be attributed to a greater redistribution of electrons around the Fermi surface at higher fluence, which in turn enhances the excitation of 5d band electrons into the 6sp conduction band in both cases.

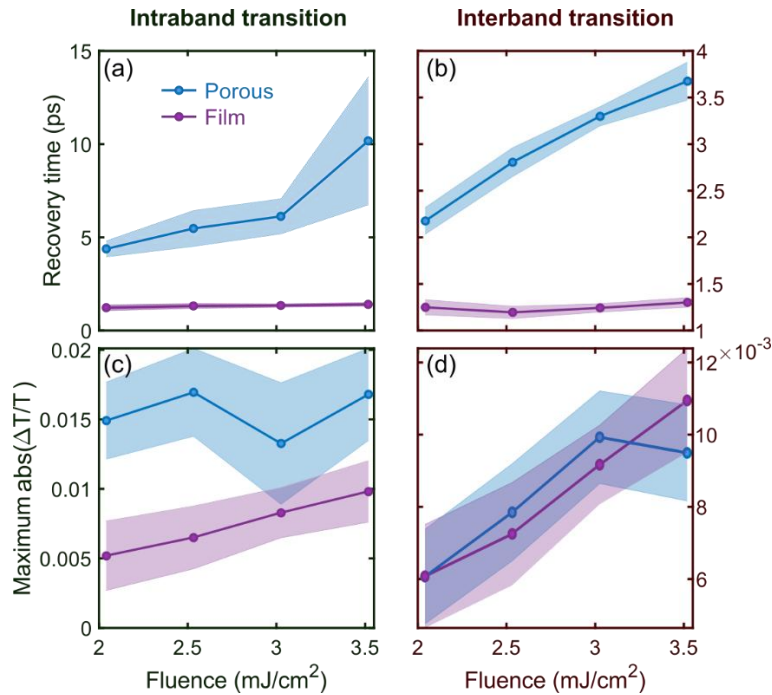

**Supplementary Figure S5. Power dependence of relaxation time and maximum transient transmission signal.** Recovery (decay) time of NPG (blue) and BG (purple) film at intraband (a) and interband (b) transitions wavelengths as a function of the pump pulse fluence. Maximum of the absolute balue of the ΔT/T signal as a function of the pump pulse fluence for NPG (blue) and BG (purple) film at intraband (c) and interband (d) transitions wavelengths.